\documentclass[sigconf]{acmart}

\usepackage{pifont}

\AtBeginDocument{%
  }

\copyrightyear{2024}
\acmYear{2024}
\setcopyright{rightsretained}
\acmConference[RecSys '24]{18th ACM Conference on Recommender Systems}{October 14--18, 2024}{Bari, Italy}
\acmBooktitle{18th ACM Conference on Recommender Systems (RecSys '24), October 14--18, 2024, Bari, Italy}\acmDOI{10.1145/3640457.3688057}
\acmISBN{979-8-4007-0505-2/24/10}

\begin{document}

\title{Bootstrapping Conditional Retrieval for User-to-Item Recommendations}

\author{Hongtao Lin}
\authornote{Both authors contributed equally to this work.}
\orcid{1234-5678-9012}
\author{Haoyu Chen}
\authornotemark[1]
\author{Jaewon Yang}
\author{Jiajing Xu}
\email{{hongtaolin, hchen, jaewonyang, jiajing}@pinterest.com}
\affiliation{%
  \institution{Pinterest}
  \city{San Francisco}
  \state{California}
  \country{USA}
}


\renewcommand{\shortauthors}{Lin et al.}

\begin{abstract}

User-to-item retrieval has been an active research area in recommendation system, and two tower models are widely adopted due to model simplicity and serving efficiency. In this work, we focus on a variant called \textit{conditional retrieval}, where we expect retrieved items to be relevant to a condition (e.g. topic). We propose a method that uses the same training data as standard two tower models but incorporates item-side information as conditions in query. This allows us to bootstrap new conditional retrieval use cases and encourages feature interactions between user and condition. Experiments show that our method can retrieve highly relevant items and outperforms standard two tower models with filters on engagement metrics. The proposed model is deployed to power a topic-based notification feed at Pinterest and led to +0.26\% weekly active users.
\end{abstract}




\maketitle

\section{Introduction}

Modern recommendation systems follow a two-stage approach: candidate retrieval and ranking. In the candidate retrieval stage, we retrieve thousands of items that a user may engage with from a corpus with billions of items. Sometimes we want the retrieved items to satisfy additional conditions (e.g. retrieve items related to a topic, filter products to a merchant), which we call \textit{conditional retrieval}.

\begin{figure}[tp]
  \centering
  \includegraphics[width=0.8\linewidth]{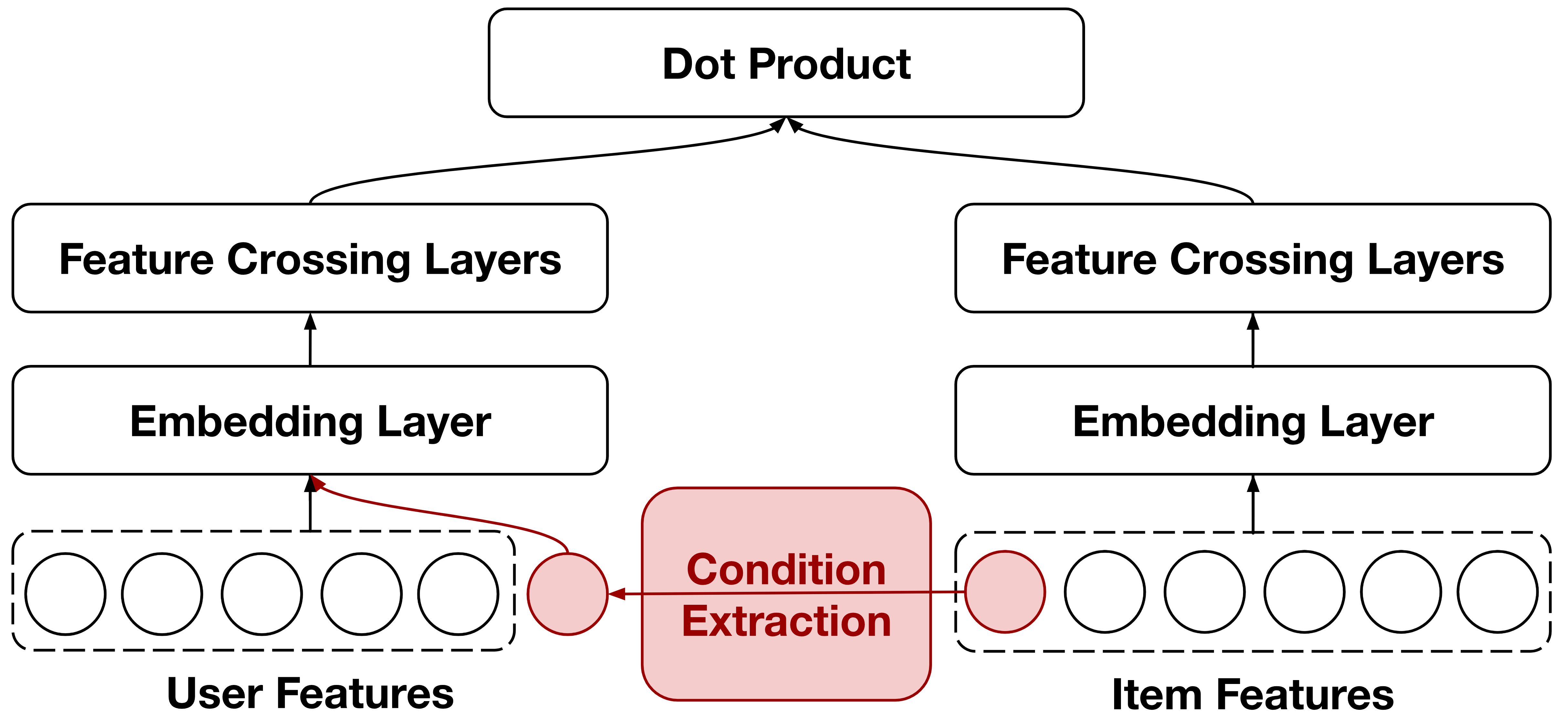}
  \caption{Two tower model architecture. The red blocks are our proposed addition for conditional retrieval model.}
  \Description{Two tower model encodes user and item features via separate neural networks (i.e. towers), which consists of an embedding layer and a few feature crossing layers. We proposed to add a conditional extraction module and conditional user tower for our work.}
  \label{fig:model}
  \vspace{-12pt}
\end{figure}

Conditional retrieval is challenging for two reasons.
Firstly, new types of condition frequently arise due to evolving ecosystems or changing business needs.
For example, in a user-to-product recommendation system, numerous attributes like color, brand, and price can serve as filter conditions. With such diversity, it is impossible to have sufficient training data tailored for each use case.
Secondly, conditional retrieval inherently involves a multi-objective optimization problem where both user engagement and condition relevance matter. In some cases, retrieving items that are highly engaging, even if they do not perfectly match the condition, can be more effective. Methods that failed to consider both objectives holistically will likely be suboptimal.

Existing candidate retrieval methods cannot address the above challenges. One commonly used method is two tower model with additional filtering. In a standard two tower model \cite{youtube,pinnerformer}, user and item features go through separate neural networks (i.e. towers) to produce embeddings (Figure~\ref{fig:model}). We can then retrieve personalized items by Approximate Nearest Neighbor (ANN) search \cite{hnsw}. To adapt to conditional retrieval use case, we either over-fetch items and post-filter based on the condition, or integrate filters during ANN search \citep{acorn}. Both methods fail to consider the relevance objective in model training and induce high serving cost.


An alternative is to utilize user engagements in such use cases and train a two tower model by providing (user, condition) in the user tower \cite{dpsr,xpert}. However, this is not feasible when bootstrapping for new use cases. For example, at Pinterest, existing notifications recommend items that a user may engage, regardless of the topic. New business needs require us to generate personalized notifications that recommend items related to a topic (i.e. topic feed).

In this paper, we propose a novel method for conditional retrieval using generic user-item engagement data. We leverage item-side features to generate artificial conditions and produce a conditional user embedding. Our method can easily adapt to new conditions, and outperforms existing methods in terms of user engagement and condition relevance.

\section{Methods} \label{sec:methods}

Standard two tower model encodes user and item features via separate neural networks (i.e. towers), which consists of an embedding layer and a few feature crossing layers (e.g. MLPs). We propose two changes to adapt to conditional retrieval (Figure \ref{fig:model}):

\textbf{Condition Extraction Module}. This module takes in the item metadata or features and generate a single condition to be used by the user tower. This module is highly dependent on the feature and condition, and is obtained independently of two tower model. For topic feed use case, we leverage an existing item-to-topic feature which maps an item to a set of topics it belongs to, and randomly sample a topic from it. 

\textbf{Conditional User Tower}. We modify the user tower to encode the condition in the embedding layer. The condition embedding is fed into feature crossing layers to allow higher order feature interactions between user and condition.

The training procedure is the same as standard two tower models. Given engaged user-item pairs, we treat them as positive examples and use other items in the same batch as negative items for the given user. The dot product between user and item embeddings is treated as the similarity score and fed into sampled softmax module \cite{mixednegative} to perform contrastive learning.

\section{Experiments}\label{sec:experiments}

At Pinterest, we regularly send users email or push notifications with personalized content recommendations. Given existing user-to-topic and item-to-topic features, we aim to introduce a new type of notification where we recommend items related to the user as well as a predefined topic. We compare the following methods for this use case:

\begin{itemize}
    \item \textbf{INDEX}. An index that maps a topic to all items according to the item-to-topic signal. We define item popularity as the number of user engagements associated with an item. During retrieval, we take top-$k$ items given a topic, sorted by item popularity.
    \item \textbf{LR}. A learned retrieval (i.e. two tower) model trained with engaged user-item pairs from the existing notification feed. 
    \item \textbf{CR}. A conditional retrieval model as described in Section \ref{sec:methods}. It uses the same training data as LR.
\end{itemize}

Both LR and CR do not guarantee perfect condition relevance. They can be combined with additional topic filters during retrieval to maximize the relevance. At Pinterest, we use in-house streaming filters \cite{streamfilter} which is similar to over-fetch and post-filter but has lower latency. It essentially performs ANN search and filter in mini-batches until we have enough items or time budget runs out.

We conducted extensive offline experiments and selected the best performing LR and CR models. We sent them to online A/B experiment which introduces the new topic-based notification. For the new notification feed, we retrieve a few hundred items via the above methods, and select around 20 items after ranking. We rank all notifications generated for a user and send a fixed number of them per day, so the overall send volume remains the same.

\begin{table}[tp]
  \caption{Online experiment results. Performance is measured by Click-Through Rate (CTR) of email and push notifications, which impact Weekly Active Users (WAU). Infra Cost refers to the annual serving cost of retrieval systems.}
  \Description{We show that conditional retrieval models outperforms other baselines in terms of engagement and infra cost.}
  \label{tab:online}
  \begin{tabular}{lccccc}
    \toprule
    Model & Filter & Email CTR & Push CTR & WAU & Infra Cost \\
    \midrule
    INDEX & - & +1.38\% & +1.25\% & +0.10\% & +40k \\
    LR & \ding{51} & +1.87\% & +2.15\% & +0.23\% & +1.2M \\
    CR & \ding{55} & +2.86\% & +2.63\% & +0.27\% & +180k \\
    CR & \ding{51} & +2.94\% & +2.58\% & +0.26\% & +300k \\
  \bottomrule
  \end{tabular}
\end{table}

As shown in Table \ref{tab:online}, all methods outperform control group which does not have this new notification type, indicating that more diverse notifications encourages users to engage more. The non-personalized INDEX approach has the lowest serving cost, but its online engagement gain is also the smallest.

Among embedding-based retrieval methods, CR achieves better performance compared to LR, showing that conditional user tower is better at capturing user engagements given condition. Interestingly, the topic filter has a neutral performance impact on CR. While the filter guarantees condition relevance, it loses the ability to retrieve items that are somewhat relevant but highly engaging. The combined effect turns out to be neutral or insignificant to user engagement.

In terms of serving cost, we could see that topic filter has a much bigger impact on LR than CR. Further analysis shows that the topic matching rate (i.e. condition relevance) of LR and CR without the filter is 20.3\% and 82.8\% respectively. This indicates that LR with topic filter has to go through a lot more candidates in ANN search phase, blowing up the serving cost. 

To sum up, CR with filter achieved the best engagement metrics with decently low serving cost. We eventually shipped it in production to guarantee condition relevance at serving time.

\section{Conclusion and Future Work}\label{sec:conclusion}

We proposed a method for conditional retrieval using the same training data as standard two tower model. Using a conditional user tower, it yields better engagement metrics and has great relevance even without applying filters. This method can be easily extended to any condition with no explicit engagement data (e.g. bootstrapping for new use cases, restricted data access due to privacy concerns).

In the future, we plan to improve the model performance by (1) including hard negatives with the same condition in sampled softmax module, (2) adding a loss between conditional user embedding and target condition to boost relevance.

\section{Speaker Bio}

\textbf{Hongtao Lin} is a Machine Learning Engineer at Pinterest since 2019. He is part of the Applied Science team and focuses on representation learning and recommendation systems. He holds a Master’s degree from University of Southern California and a Bachelor’s degree from Shanghai Jiao Tong University.

\begin{acks}
This work cannot be accomplished without the help from Prabhat Agarwal, Nikil Pancha, Rui Liu, Yuxiang Wang and Bowen Deng. We would like to thank them for their support and contributions along the way.
\end{acks}

\bibliographystyle{ACM-Reference-Format}
\bibliography{sample-base}


\begin{thebibliography}{8}


\ifx \showCODEN    \undefined \def \showCODEN     #1{\unskip}     \fi
\ifx \showDOI      \undefined \def \showDOI       #1{#1}\fi
\ifx \showISBNx    \undefined \def \showISBNx     #1{\unskip}     \fi
\ifx \showISBNxiii \undefined \def \showISBNxiii  #1{\unskip}     \fi
\ifx \showISSN     \undefined \def \showISSN      #1{\unskip}     \fi
\ifx \showLCCN     \undefined \def \showLCCN      #1{\unskip}     \fi
\ifx \shownote     \undefined \def \shownote      #1{#1}          \fi
\ifx \showarticletitle \undefined \def \showarticletitle #1{#1}   \fi
\ifx \showURL      \undefined \def \showURL       {\relax}        \fi
\providecommand\bibfield[2]{#2}
\providecommand\bibinfo[2]{#2}
\providecommand\natexlab[1]{#1}
\providecommand\showeprint[2][]{arXiv:#2}

\bibitem[Covington et~al\mbox{.}(2016)]%
        {youtube}
\bibfield{author}{\bibinfo{person}{Paul Covington}, \bibinfo{person}{Jay Adams}, {and} \bibinfo{person}{Emre Sargin}.} \bibinfo{year}{2016}\natexlab{}.
\newblock \showarticletitle{Deep neural networks for youtube recommendations}. In \bibinfo{booktitle}{\emph{Proceedings of the 10th ACM conference on recommender systems}}. \bibinfo{pages}{191--198}.
\newblock


\bibitem[Koh and Wu(2022)]%
        {streamfilter}
\bibfield{author}{\bibinfo{person}{Tim Koh} {and} \bibinfo{person}{George Wu}.} \bibinfo{year}{2022}\natexlab{}.
\newblock \showarticletitle{{Manas HNSW Streaming Filters}}.
\newblock  (\bibinfo{year}{2022}).
\newblock
\urldef\tempurl%
\url{https://medium.com/pinterest-engineering/manas-hnsw-streaming-filters-351adf9ac1c4}
\showURL{%
\tempurl}


\bibitem[Malkov and Yashunin(2018)]%
        {hnsw}
\bibfield{author}{\bibinfo{person}{Yu~A Malkov} {and} \bibinfo{person}{Dmitry~A Yashunin}.} \bibinfo{year}{2018}\natexlab{}.
\newblock \showarticletitle{Efficient and robust approximate nearest neighbor search using hierarchical navigable small world graphs}.
\newblock \bibinfo{journal}{\emph{IEEE transactions on pattern analysis and machine intelligence}} \bibinfo{volume}{42}, \bibinfo{number}{4} (\bibinfo{year}{2018}), \bibinfo{pages}{824--836}.
\newblock


\bibitem[Pancha et~al\mbox{.}(2022)]%
        {pinnerformer}
\bibfield{author}{\bibinfo{person}{Nikil Pancha}, \bibinfo{person}{Andrew Zhai}, \bibinfo{person}{Jure Leskovec}, {and} \bibinfo{person}{Charles Rosenberg}.} \bibinfo{year}{2022}\natexlab{}.
\newblock \showarticletitle{Pinnerformer: Sequence modeling for user representation at pinterest}. In \bibinfo{booktitle}{\emph{Proceedings of the 28th ACM SIGKDD conference on knowledge discovery and data mining}}. \bibinfo{pages}{3702--3712}.
\newblock


\bibitem[Patel et~al\mbox{.}(2024)]%
        {acorn}
\bibfield{author}{\bibinfo{person}{Liana Patel}, \bibinfo{person}{Peter Kraft}, \bibinfo{person}{Carlos Guestrin}, {and} \bibinfo{person}{Matei Zaharia}.} \bibinfo{year}{2024}\natexlab{}.
\newblock \showarticletitle{ACORN: Performant and Predicate-Agnostic Search Over Vector Embeddings and Structured Data}.
\newblock \bibinfo{journal}{\emph{arXiv preprint arXiv:2403.04871}} (\bibinfo{year}{2024}).
\newblock


\bibitem[Vemuri et~al\mbox{.}(2023)]%
        {xpert}
\bibfield{author}{\bibinfo{person}{Hemanth Vemuri}, \bibinfo{person}{Sheshansh Agrawal}, \bibinfo{person}{Shivam Mittal}, \bibinfo{person}{Deepak Saini}, \bibinfo{person}{Akshay Soni}, \bibinfo{person}{Abhinav~V Sambasivan}, \bibinfo{person}{Wenhao Lu}, \bibinfo{person}{Yajun Wang}, \bibinfo{person}{Mehul Parsana}, \bibinfo{person}{Purushottam Kar}, {et~al\mbox{.}}} \bibinfo{year}{2023}\natexlab{}.
\newblock \showarticletitle{Personalized Retrieval over Millions of Items}. In \bibinfo{booktitle}{\emph{Proceedings of the 46th International ACM SIGIR Conference on Research and Development in Information Retrieval}}. \bibinfo{pages}{1014--1022}.
\newblock


\bibitem[Yang et~al\mbox{.}(2020)]%
        {mixednegative}
\bibfield{author}{\bibinfo{person}{Ji Yang}, \bibinfo{person}{Xinyang Yi}, \bibinfo{person}{Derek Zhiyuan~Cheng}, \bibinfo{person}{Lichan Hong}, \bibinfo{person}{Yang Li}, \bibinfo{person}{Simon Xiaoming~Wang}, \bibinfo{person}{Taibai Xu}, {and} \bibinfo{person}{Ed~H Chi}.} \bibinfo{year}{2020}\natexlab{}.
\newblock \showarticletitle{Mixed negative sampling for learning two-tower neural networks in recommendations}. In \bibinfo{booktitle}{\emph{Companion proceedings of the web conference 2020}}. \bibinfo{pages}{441--447}.
\newblock


\bibitem[Zhang et~al\mbox{.}(2020)]%
        {dpsr}
\bibfield{author}{\bibinfo{person}{Han Zhang}, \bibinfo{person}{Songlin Wang}, \bibinfo{person}{Kang Zhang}, \bibinfo{person}{Zhiling Tang}, \bibinfo{person}{Yunjiang Jiang}, \bibinfo{person}{Yun Xiao}, \bibinfo{person}{Weipeng Yan}, {and} \bibinfo{person}{Wen-Yun Yang}.} \bibinfo{year}{2020}\natexlab{}.
\newblock \showarticletitle{Towards personalized and semantic retrieval: An end-to-end solution for e-commerce search via embedding learning}. In \bibinfo{booktitle}{\emph{Proceedings of the 43rd International ACM SIGIR Conference on Research and Development in Information Retrieval}}. \bibinfo{pages}{2407--2416}.
\newblock


\end{thebibliography}

\end{document}